\def\year{2016}\relax
\documentclass[letterpaper]{article}
\usepackage{aaai16}
\usepackage{times}
\usepackage{helvet}
\usepackage{courier}
\usepackage{amssymb}
\usepackage{amsmath}
\usepackage{subfigure}
\usepackage[english]{babel}
\usepackage{amsmath}
\usepackage{graphicx}
\usepackage{array}
\usepackage{multirow}
\usepackage{booktabs}
\usepackage{upquote}
\usepackage{float}
\usepackage[ruled]{algorithm2e}
\usepackage{url}
\SetKwRepeat{Do}{do}{while}
\begin{document}
%
\title{Predicting Online Protest Participation of Social Media Users}

\author{
Suhas Ranganath \\ srangan8@asu.edu \\Arizona State University \And Fred Morstatter \\Fred.Morstatter@asu.edu \\ Arizona State University \And Xia Hu\\hu@cse.tamu.edu \\Texas A\&M University \AND Jiliang Tang \\jlt@yahoo-inc.com\\Yahoo Labs \And Suhang Wang\\ swang187@asu.edu \\Arizona State University \And Huan Liu\\ Huan.Liu@asu.edu\\Arizona State University\\
}

\maketitle
\begin{abstract}
Social media has emerged to be a popular platform for people to express their viewpoints on political protests like the Arab Spring. Millions of people use social media to communicate and mobilize their viewpoints on protests. Hence, it is a valuable tool for organizing social movements. However, the mechanisms by which protest affects the population is not known, making it difficult to estimate the number of protestors. In this paper, we are inspired by sociological theories of protest participation and propose a framework to predict from the user's past status messages and interactions whether the next post of the user will be a declaration of protest. Drawing concepts from these theories, we model the interplay between the user's status messages and messages interacting with him over time and predict whether the next post of the user will be a declaration of protest. We evaluate the framework using data from the social media platform Twitter on protests during the recent Nigerian elections and demonstrate that it can effectively predict whether the next post of a user is a declaration of protest.
\end{abstract}

Social media has emerged as a popular information and communication channel for protest-related issues \cite{IAAI159652,ContractorCMSF15,tufekci2012social}. It provides an open and accessible platform for people to put forth views on issues affecting them. Millions of people, therefore, use social media to declare protest, mobilize opinion and participate in discussions on these issues. For example, electoral malpractice was suspected during the recent elections in Nigeria \cite{knifeedge}, and users employed social media to express and mobilize viewpoints.

We concentrate on posting behavior of users and define online protest participation as the act of declaring protest through status messages. However, it is a nontrivial problem to directly predict online protest participation of social media users. Millions of users post in social media during popular social movements, and protest organizers have to go through them to identify potential participants. The mechanisms by which the protest affects the population cannot be fully observed \cite{lin2013bigbirds}, and protest organizers cannot easily estimate the number of protestors. Designing algorithms to predict whether the next post of a user will be a declaration of protest will enable protest organizers to anticipate the behavior of the protestors and estimate the total number of participants.

This task faces several challenges. First, the ways in which protest-related events affect a person are not observable, resulting in a lack of knowledge of factors operating at that time causing his next post to be a declaration of protest. Second, a user is subject to various types of influence in his past, and many of them are in conflict with each other. This may lead to ambiguities on whether his posts will contain declarations of protest in the future. Finally, each user can post a large amount of content and interact with many people, leading to issues of scalability.

Sociological studies have theorized factors from an individual's history causing his next post to be a declaration of protest. A user will be more likely to protest if his social ties have reached out to him with protest-related messages in his past \cite{schussman2005process}. The chances of him protesting are bolstered if these messages are sent by people interested in protest related issues \cite{lim2008social}. The likelihood of protest is reduced if people uninterested in protest-related issues have reached out to him in the past with messages unrelated to protests. \cite{snow1980social}.

Inspired by these sociological theories, we utilize the user's previous status messages and messages interacting with him to predict whether his next post will be a declaration of protest. To model the effect of the interactions on the user's status messages, we build upon concepts from the Brownian motion theory of fluid particle motion \cite{zhou2003network}. This theory models the path of fluid particles as other fluid particles come in contact with them over time. We draw analogies from these concepts to model the probability of the user declaring protest as other users reach out to him over time. The primary contributions of this work are:
\begin{itemize}
\item Formally defining the problem of predicting whether the next post of a user will be a declaration of protest based on past status messages and messages interacting with him,
\item Demonstrating the applicability of sociological theories of protest participation in online social media data,
\item Proposing a framework that predicts online protest participation of a user, and
\item Evaluating the framework using a real world dataset from Twitter on the Nigerian election protest in March 2015.
\end{itemize}

\begin{figure}
\begin{center}
\includegraphics[scale=0.20]{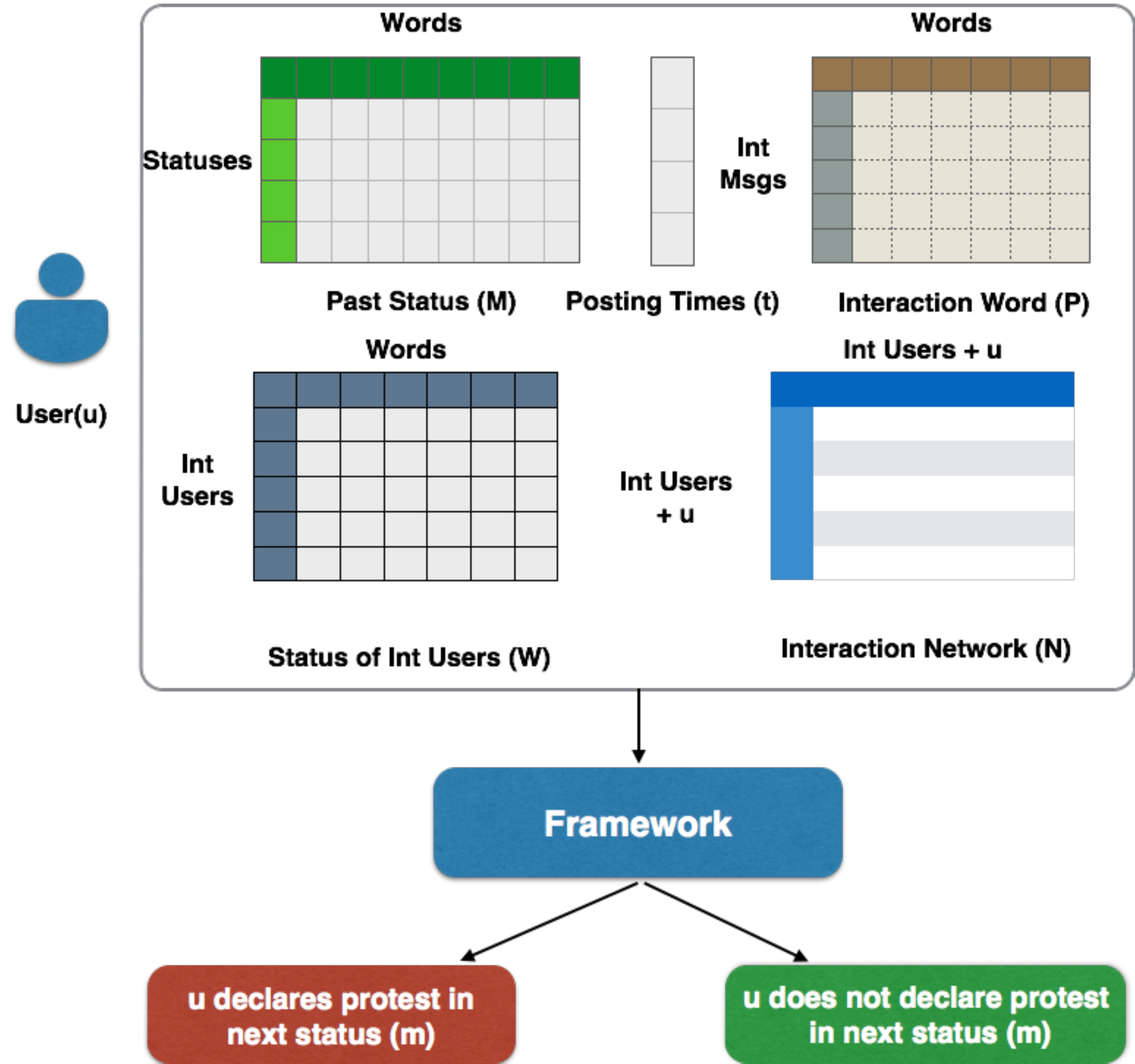}
\end{center}
\caption{The proposed framework to predict if the next post of user $u$ is a declaration of protest}
\label{fig:framework}
\end{figure}

\section{Problem Statement}
\label{sec:prosta}
In this section, we present the notations and formally state the problem. Boldface uppercase letters (e.g., $\mathbf{X}$) denote matrices and lowercase letters (e.g., $\mathbf{x}$) denote vectors. $\mathbf{X}_{ij}$ signifies the element in the $i^{th}$ row and $j^{th}$ column of $\mathbf{X}$. $\mathbf{x} _i$ denotes the $i^{th}$ element of $\mathbf{x}$. The $i^{th}$ column of a matrix $\mathbf{X}$ is denoted by $\mathbf{X}_i$ and its $i^{th}$ row by $\mathbf{X}_{i'}$. The Frebonius norm of a matrix $\mathbf{X}$ is denoted as $||\mathbf{X}||_{F}=\sqrt{\sum_{i,j}{\mathbf{X}_{ij}^2}}$.

The terms related to proposed framework are illustrated in Fig \ref{fig:framework}. Let the set of candidate users be denoted as $\mathcal{U}$. For each user $u \in \mathcal{U}$, let $\mathbf{t}$ contain the posting times of a set of status messages denoted by $\mathcal{M}$. We construct a dictionary of words in the status messages of candidate users, messages that reach out to them and status messages of users posting them. Previous status messages of candidate users are contained in the matrix $\mathbf{M} \in \mathbb{R}^{l \times m}$. Here $l$ is the number of status messages, and $m$ is the total number of words in them. Messages reaching out to $u$ are contained in the matrix $\mathbf{P} \in \mathbb{R}^{o \times r}$. Here $o$ is the total number of such posts and $r$ is the total number of words in them. Similarly, the status messages of users posting such messages are contained in $\mathbf{W} \in \mathbb{R}^ {o \times q}$. Here $q$ is the total number of words in the status messages. Let the user $u$ and users who post messages interacting with them constitute the set $\mathcal{G}$. $\mathbf{N} \in \mathbb{R}^{d \times d}$ denotes the interaction network between these users, where $d$ is the total number of users, and
\begin{equation*}
\mathbf{N}_{ij} =
\begin{cases}
t & g_i \text{ interacts with } g_j \text{ } t \text{ times}; g_i, g_j \in \mathcal{G} ,\\
0 & \text{otherwise}
\end{cases}
\end{equation*}

The problem can be formally defined as \textit{``Given a user u, his posting time vector $\mathbf{t}$, his status message matrix $\mathbf{M}$, interaction word matrix $\mathbf{P}$, the word matrix of users posting these interactions $\mathbf{W}$ and the interaction network $\mathbf{N}$, predict if the next post of user $u$ will be a declaration of protest.''}

\section{Data Analysis}
\label{sec:datasets}

In this section, we describe the data collection methods and present statistics related to the collected dataset. We then propose postulates from sociological theories of protest participation that form the conceptual basis of the framework and utilize the datasets to verify them.

The data is from the Nigerian general election that took place on March 2015. This event took place amid insurgencies of Boko Haram \cite{bokoharam} and discrepancies in the voting process \cite{knifeedge}. We collected messages from the Twitter Streaming API geotagged within Nigeria using \cite{kumar2011tweettracker}. Assisted by experts, we prepared a set of keywords and hashtags denoting declarations of protests to filter out tweets that did not contain them. The earliest tweet in our dataset was posted on Feb. 25, 2015, and the last tweet on Apr. 27, 2015. We collected a total of 2,686 posts potentially containing declarations of protest by users.

Posts containing protest-related hashtags and keywords do not necessarily mean it is declaring protest. For example the post ``That same guy you can't ever resist.'' contains the protest-related keyword ``resist'', but is not a declaration of protest. To address this discrepancy, we used Amazon Mechanical Turk. Each post was given to the workers along with instructions to determine if the user is declaring protest in it. Three distinct workers evaluated each post, and we employed majority voting to choose the category of the post. In total, the workers categorized 626 posts where the user expressed protest, which we refer to as positive labels. The workers categorized that 2,060 posts did not contain a declaration of protest, which we refer to as negative labels. We use these posts as ground truth for future experiments and users who post them form the set $\mathcal{U}$.

The Twitter Streaming API shows a bias in keyword and geo-tagged distributions \cite{Morst-etal13}, so the data might not represent the actual spread of protests. To overcome this, we treat each user as a separate entity and use his historical information for prediction. For each user $u \in \mathcal{U}$, we collect their status messages, a maximum of 200 posts, and extract the set of users he mentioned to construct the set $\mathcal{G}$. As we are interested in predicting whether the post of $u$ at time t is a declaration of protest, we do not explicitly denote if his previous posts are a declaration of protest. We collect the status messages of the users in $\mathcal{G}$ and retain only those users who have mentioned $u$ back. We then construct the matrices $\mathbf{M}$, $\mathbf{W}$, $\mathbf{P}$ and $\mathbf{N}$. In total, the dataset contains 63,983 users, around 11 million interactions and 105 million posts by users who mentioned the candidates.

Our first postulate is regarding the ability of appropriate social ties that can enhance protest participation \cite{lim2008social}. It can be stated that ``if the user $u$ is previously mentioned in a protest related message by people interested in it, the next post of the user is likely to be a declaration of protest''. We first denote the set of words $\mathcal{L}$ in the positive labels by the crowdsourced workers and denote them as protest-related words. We then select columns corresponding to these words from $\mathbf{P}$ and $\mathbf{W}$ and denote them as $\mathbf{A}$ and $\mathbf{B}$ respectively. To measure the extent that user is mentioned in protest-related messages by people interested in them, we compute $\sum_{i=1}^{P}\mathbf{A}_{i'}{\mathbf{B}_{i'}}^T$, where $P$ is the total number of messages. The higher the number of protest-related messages by interested users mentioning $u$ in his past, the higher will be this quantity. We repeat this for all the users with positive labels and a similar number of users with negative labels. A paired t-test shows that positively labeled users have a significantly higher score with $p<0.01$ than negatively labeled users.

Our next postulate is regarding the ties to alternative networks that can hinder protest participation \cite{snow1980social}. It can be stated as ``the more the user is mentioned in a message unrelated to the protest by people not interested in it in his past, the lower the chance of his next post being a declaration of protest''. We select the columns from $\mathbf{P}$ and $\mathbf{W}$ corresponding to words not present in $\mathcal{L}$ and denote the matrices as $\mathbf{C}$ and $\mathbf{D}$. To measure the extent that a user is mentioned in matters unrelated to protests by people not interested in protests, we compute $\sum_{i=1}^{P}\mathbf{C}_{i'}{\mathbf{D}_{i'}}^T$. The greater the number of unrelated messages $u$ is mentioned in by uninterested users in his past, the higher will be this sum. We repeat this for all the users given positive labels and a similar number of users given negative labels. A paired t-test shows that negatively labeled users have a significantly higher score than positively labeled users with $p<0.05$.

\begin{table}
\begin{center}
\begin{tabular}{l r}
\toprule
\textbf{Parameter} & \textbf{Value}\\
\midrule
Candidate Posts & 2,686 \\
Protest Posts & 626\\
Non Protest Posts & 2,060\\
Candidate Users & 2,686 \\
Previous Posts of Candidate Users & 362,485 \\
Interacted Users & 63,983\\
Interactions & 11,976,235\\
Posts of Interacted Users & 105,513,184\\
\bottomrule
\end{tabular}
\end{center}
\caption{Dataset of protests during Nigerian elections.}
\label{tab:dataset}
\end{table}

\section{The Proposed Framework}

In this section, we build upon the presented postulates to design a computational framework that predicts whether the next post of a user will be a declaration of protest. It accomplishes this by modeling the interplay between status messages and messages interacting with him in his past.

The matrices in Fig \ref{fig:framework} have a large number of elements, making scalability an important issue. To improve scalability, we represent them in a shared latent dimensional space. Latent representations also give a semantic meaning to the interactions and the status messages, as each dimension can be interpreted as a cluster. From the first postulate, the probability of $u$ posting in dimension $i$ increases if a user interested in dimension $i$ mentioned him in the past in a post related to $i$. From the second postulate, the probability decreases if he is mentioned in a post unrelated to dimension $i$ in his past by a user with low interests in $i$. Posts related to other dimensions might affect his probability of posting in $i$. For instance, a post in the dimension related to politics might bolster the likelihood of the next post declaring protest, and a post associated with fashion might diminish it. We model these concepts to develop a framework that predicts whether the next post of the user $u$ will be a declaration of protest.

We first construct a latent dimensional representation for the matrices $\mathbf{P}$ and $\mathbf{W}$ from Fig \ref{fig:framework}. The latent dimension representation of message $k$ mentioning the user is given by $\mathbf{P}_{k'}\mathbf{U}=\mathbf{L}_{k'}$ and the content of the users posting those messages by $\mathbf{W}_{k'}\mathbf{V}=\mathbf{X}_{k'}$. Here $\mathbf{U} \in \mathbb{R}^{r \times I}$ and $\mathbf{V} \in \mathbb{R}^{q \times I}$ are latent dimension representations of the content of messages and users posting them respectively, with $I$ dimensions.

Let $\mathbf{s}_i^M$ and $\mathbf{s}_i^{M-1}$ give the predicted latent dimension representation of the $M^{th}$ and $M-1^{th}$ previous status messages of $u$ in dimension $i$, ordered chronologically from the earliest post. Let $k$ be a message mentioning the user $u$ posted between his $M^{th}$ and $M-1^{th}$ previous status message. The change in $\mathbf{s}_i^M$ due to $k$ can be modeled as $\mathbf{L}_{ki}\mathbf{T}_{i'}{\mathbf{X}_{k'}}^T$, where $\mathbf{T}$ is the dimension correlation matrix. This quantity has a high value when a message related to dimension $i$ is posted by a person with high interests in $i$ mentions the user, modeling the first postulate. It has a low value when a message unrelated to dimension $i$ is posted by a user with low interests in $i$, thus modeling the second postulate. It also increases or decreases when a message related to dimension $j$ is posted by a person interested in $j$ and dimension $j$ has a high positive or negative correlation with $i$ respectively. We represent change in $\mathbf{s}_i^{M}$ due to all posts mentioning $u$ between the $M^{th}$ and $M-1^{th}$ previous status message in the form of a differential equation.
\begin{equation}
\frac{d\mathbf{s}_i^M}{\mathbf{s}_i^M}=\mu_i(1+ \sum_{k \in \mathcal{P}_M} \mathbf{L}_{ki}\mathbf{T}_{i'}{\mathbf{X}_{k'}}^T)dt + \sigma_i dW.
\label{eqn:diff}
\end{equation}
Here $\mu$ is the default change of $\mathbf{s}_i^M$ when there are no messages interacting with him. $\mathcal{P}_M$ is the total set of messages aggregated for model simplicity. $W$ is the Wiener process to account stochastic variations. \cite{oksendal2003stochastic}. Let $\mathbf{n}_i=1+\sum_{k \in \mathcal{P}_M} \mathbf{L}_{ki}\mathbf{T}_{i'}{\mathbf{X}_{k'}}^T$.

The intuition for Eq. \ref{eqn:diff} can be drawn from Geometric Brownian Motion (GBM) \cite{karatzas2012brownian} which models the path of fluid particles as they interact with other particles over time. It models two quantities; drift, to account for the deterministic effects and volatility, for the unpredictable effects of the interactions. In Eq. \ref{eqn:diff}, $\mu_i \mathbf{n}_i$ is equivalent to the drift, and $\sigma_i$ to volatility. The differential equation has the following solution
\begin{equation}
\mathbf{s}_i^M=\mathbf{s}^{M-1}_i exp \Big(\mathbf{\mu}_i (\mathbf{n}_i - \frac{\sigma_i^2}{2})t + \sigma_i W \Big),
\label{eqn:update}
\end{equation}
where $t$ is the time elapsed from when the $M-1^{th}$ status message was posted.

We next present a method to optimally learn the parameters $\mathcal{K}=\{\mu,\sigma,\mathbf{U},\mathbf{V},\mathbf{T}\}$. From the properties of GBM \cite{karatzas2012brownian}, $ln(\mathbf{s}_i^{M+1})\sim \mathcal{N} (\mathbf{m}_i,\mathbf{v}_i)$. Here $\mathbf{m}_i=ln(\mathbf{s})^{M-1}_i + \big(\mathbf{\mu}_i \mathbf{n}_i -\frac{\sigma_i^2}{2}\big)t$ is the mean and $\mathbf{v}_i=\sigma_i \sqrt{t}$ is the variance of the Gaussian Distribution. We denote the predicted value of the $M+1^{th}$ status message as $ln(\mathbf{s}_i^M)$ for all $i$ and equate it to the latent representation of the actual message. Let the word vector of the $M^{th}$ status message be $\mathbf{g}$ and its latent representation be $\mathbf{g}\mathbf{V}=\mathbf{a}$. We set $ln(\mathbf{s}_i^M) \approxeq \mathbf{a}_i$ for all $i$. The likelihood function then becomes
\begin{equation}
\mathcal{L}(\theta)=\frac{1}{\sigma_i \sqrt{2 \pi t}} exp \Big(- \frac{(\mathbf{a}_i -\mathbf{m}_i)^2}{2 (\mathbf{v}_i)^2} \Big)
\label{eqn:likelihood}
\end{equation}

\begin{algorithm}
\KwData{$\mathcal{U}$, $\mathcal{M}$, $\mathcal{K}$, $w_{reg}$}
\KwResult{Labels of Protest Declarations}
\For{$u \in \mathcal{U}$}{
Randomly initialize $\mathcal{K}$\;
\For{ $M \in \mathcal{M}$ }{
\For{ all latent dimensions $i$ }{
Compute $\mathcal{L}(\theta)$ from Eq. \ref{eqn:likelihood} and $f_r$ from Eq. \ref{eqn:socreg}\;
Compute $f_t=-ln(\mathcal{L}(\theta)) + w_{reg}f_r$ \;
\Do{$f_t$ does not converge}{
Update values of $\mathcal{K}$ using Eq. \ref{eqn:update1} to \ref{eqn:update4}.\;
Compute $\mathbf{m}_i$, $\mathbf{v}_i$ from updated values.\;
Update values of $\mathcal{L}(\theta)$ , $f_r$ and $f_t$.\;
}
Draw $ln({\mathbf{s}}_i^{M+1}) \sim \mathcal{N} (\mathbf{m}_i, \mathbf{v}_i)$\;
}
}
Predict feature vector $\mathbf{u}$ as $ln(\mathbf{s}_i^{size(\mathcal{M})+1}) \forall i$ \;
}
Construct $\mathbf{F}$ from $\mathbf{u}$ for all $u \in \mathcal{U}$ \;
Classify the feature set $\mathbf{F}$ to obtain labels\;
\caption{Predicting Online Protest Participation}
\end{algorithm}

Drawing from the principles of network homogeneity \cite{marsden1988homogeneity}, the latent dimensions of users can be similar to the users they interact with. We model this by introducing the regularization function
\begin{equation}
f_r= \sum_{\mathbf{N}_{ij}^M=1} (sig(\mathbf{X}_{i'}{\mathbf{X}_{j'}}^T)-1 )^2,
\label{eqn:socreg}
\end{equation}
where $sig(x)=\frac{1}{1+e^{-x}}$ denotes the sigma function, and $\mathbf{N}^M$ is the interaction network of user $u$ at the time of the $M^{th}$ status message. The overall optimization function is
\begin{equation}
\min_{\mu,\sigma,\mathbf{U},\mathbf{V},\mathbf{T}}-ln(\mathcal{L}(\theta)) + w_{reg}f_r,
\label{eqn:optfunc}
\end{equation}
where $w_{reg}$ is the contribution of the regularizer. Letting $\mathbf{a}_i -\mathbf{m}_i=\mathbf{c}_i$, the update equations for each  $i$ are
\begin{equation*}
\mu_{i} \leftarrow \mu_{i} + \eta \frac{1}{\sigma_i^2}\mathbf{n}_i\mathbf{c}_i, \sigma_i \leftarrow \sigma_i + \eta \frac{1}{t\sigma_i^3}\big(\mathbf{c}_i(-\mathbf{c}_i +t \sigma_i^2)+t \sigma_i^2\big)
\label{eqn:update1}
\end{equation*}
\begin{equation*}
\mathbf{U}_i \leftarrow \mathbf{U}_i +\eta \frac{\mu_i}{\sigma^2_i}\mathbf{c}_i\sum_{k \in \mathcal{P}_M}{\mathbf{P}_{k'}}^T\mathbf{X}_{k'}{\mathbf{T}_{i'}}^T,
\label{eqn:update2}
\end{equation*}

\begin{equation*}
\mathbf{T}_{i'} \leftarrow \mathbf{T}_{i'} +\eta \frac{\mu_i}{\sigma^2_i}\mathbf{c}_i\sum_{k \in \mathcal{P}_M}\mathbf{L}_{ki}^T\mathbf{X}_{k'}
\label{eqn:update3}
\end{equation*}
\small
\begin{equation}
\mathbf{V}\leftarrow \mathbf{V}+\eta\bigg(\frac{2}{\sigma^2_it}\mathbf{c}_i\big(-\mathbf{m}_i+\mu_i t \sum_{k \in \mathcal{P}_M}{\mathbf{W}_{k'}}^T\mathbf{L}_{ki}\mathbf{T}_{i'} \big)+ w_{reg} \mathbf{Y}\bigg).
\label{eqn:update4}
\end{equation}
\normalsize
Here
\begin{equation}
\mathbf{Y}=\sum_{\mathbf{N}_{ij}^M=1}2sig(g_{ij})(sig(g_{ij})-1)^2{\mathbf{X}_{i'}}^T\mathbf{X}_{j'}\mathbf{V},
\end{equation}
where $\mathbf{X}_i^T\mathbf{X}_j=g_{ij}$. We repeat this all for the status messages of $u$, updating values of $\mathbf{m}$ and $\mathbf{v}$ at every instance.

After learning the parameters, we predict the latent membership of the next post of $u$ as $\mathbf{u}_i \sim {\mathcal{N} (\mathbf{m}_i,\mathbf{v}_i)}$ for all $i$. We repeat this for all the users in the dataset to build a feature set $\mathbf{F}$, whose each row corresponds to the latent dimension $\mathbf{u}$ of a single user. We then classify $\mathbf{F}$ to decide whether the next post of the user is a declaration of protest. The algorithm is scalable due to the use of latent dimensions and can be used in large datasets usually encountered in social media.

\section{Experiments}
We evaluate the framework using a real world dataset presented in Table \ref{tab:dataset} by answering the following questions: How does the framework perform in determining whether the next post of the user is a declaration of protest? What is the effect of the varying proportions of the interaction network information on the performance of the framework? How does the framework perform for different proportions of training data? We first present the designed experimental settings and then answer each of these questions in detail.

\subsection{Experimental Settings}
We now present the metrics and baselines used for evaluating the algorithm. In our dataset, there is an imbalance of positive and negative examples with a ratio of about 1:6. Therefore, we choose metrics that are capable of dealing with data imbalance. We use three metrics: accuracy, AUC, and the F1 measure computed for the positive class. The following baselines are employed as performance benchmarks.
\begin{itemize}
\item \textbf{Random Label Assignment}: Whether the user is protesting or not protesting is randomly assigned. This is repeated for 100 trials, and the mean value is presented.
\item \textbf{Info Prop} \cite{jin2014modeling}: The authors model declarations of protest as it spreads through the information propagation network with external influence.
\item \textbf{Topic} \cite{godin2013using}: This work predicts the topics of the next post of the user from the topics of his previous posts. We present the predicted topics to classify whether the next post of the user is a declaration of protest.
\item \textbf{Sim Net} \cite{kywe2012recommending}: The authors use collaborative filtering methods to predict features of the next post of the user. They incorporate the content of similar users and posts while computing the features.
\item \textbf{Time Topic} \cite{ma2014tagging}: The paper predicts the topics of the next post of the user by integrating the past content of the user, content at the time of the candidate message and interaction network between candidate users.
\item \textbf{Our Method -Int}: We implement $\textbf{Algorithm 1}$ with only the past status messages of the user i.e. $\mathbf{n}_i=\mu_i$. This is to assess the value of user interactions in predicting whether the next post of the user is a declaration of protest.
\end{itemize}

\subsection{Performance Evaluation}

We now evaluate the performance of the algorithm using Accuracy, AUC, and F1 measure, and compare it with the baselines. In this experiment, we set the value of information network parameter at $w_{reg}=0.1$ and the number of latent dimensions $I=50$. We set the earliest 50\% of the candidate posts for training and the rest for testing. Later in the section, we will experiment with different values of $w_{reg}$ and training data size. We use a linear discriminant classifier on the feature set $\mathbf{F}$ and illustrate the results in Table \ref{tab:expresults}.

From Table \ref{tab:expresults}, we see that random assignment performs poorly in all the three metrics showing the difficulty of the problem. \cite{jin2014modeling} performs slightly better than random assignment, showing that diffusion-based algorithms might not work in cases where the entire propagation network is not available. \cite{godin2013using} gives better performance by building topic models from the user's previous posts, showing the utility of harnessing the user's history to predict his participation.

The baseline \cite{kywe2012recommending} utilizes collaborative filtering and gives a similar performance as \cite{godin2013using}, even though it uses word features rather than topic models. The similar performance showcases that using relational attributes of users assists in enhancing the performance. This is further illustrated by the improvement over previous baselines shown by \cite{ma2014tagging} which integrates content, network and temporal information to predict characteristics of the user's next post.

The improvement by \textbf{Our Method -Int} demonstrates that the framework effectively models the user's past status messages using concepts from GBM for the task. The significant improvement in \textbf{Our Method} demonstrates that sociological theories of protest participation to model the interplay between interactions and user content are useful in determining whether the next post of a user will be a declaration of protest. It also demonstrates the effectiveness of the framework for modeling these theories. A paired t-test showed that the method improves significantly over the baselines.

In summary, the proposed framework significantly outperforms the baselines and is useful for predicting whether the next post of a user will be a declaration of protest. In the next section, we will examine the variation of the performance with different values of the interaction network parameter.

\begin{table}
\begin{center}
\begin{tabular}{l r r r}
\toprule
\textbf{Method} & \textbf{Accuracy} & \textbf{AUC} & \textbf{F1} \\
\midrule
Random &0.491&0.116& 0.607\\
Info \cite {jin2014modeling} &0.531 & 0.120 & 0.629 \\
Topic \cite{godin2013using} &0.587&0.133&0.692\\
Sim \cite{kywe2012recommending} &0.570 & 0.147 & 0.693 \\
Time \cite{ma2014tagging} & 0.599 &0.150&0.715\\
Our Method -Int & 0.636 & 0.177 & 0.765\\
Our Method &\textbf{0.702}&\textbf{0.204}&\textbf{0.817}\\
\bottomrule
\end{tabular}
\end{center}
\caption{Performance of the proposed framework.}
\label{tab:expresults}
\end{table}

\begin{table*}
\begin{center}
\begin{tabular}{cc|c|c|c|c|c|c|c|c|c|c}
\cline{3-10}
& & \multicolumn{8}{ c| }{$w_{reg}$} \\ \cline{3-10}
& & \textbf{0} & \textbf{0.1} & \textbf{0.2} & \textbf{0.5} & \textbf{1} & \textbf{2} & \textbf{5} & \textbf{10} \\ \cline{1-10}
\multicolumn{1}{ |c }{\multirow{3}{*}{Metrics} } &
\multicolumn{1}{ |c| }{\textbf{Accuracy}} &0.681&0.702&0.697&\textbf{0.716}&0.676&0.691&0.676&0.689 \\ \cline{2-10}
\multicolumn{1}{ |c }{} &
\multicolumn{1}{ |c| }{\textbf{AUC}} &0.192&0.204&0.201&\textbf{0.213}&0.189&0.191&0.189&0.195 \\ \cline{2-10}
\multicolumn{1}{ |c }{} &
\multicolumn{1}{ |c| }{\textbf{F1 Measure}} & 0.798&0.817&0.813&\textbf{0.828}&0.794&0.810&0.794&0.807 \\ \cline{1-10}
\end{tabular}
\end{center}
\caption{Performance of the framework with varying values of information network parameter}
\label{tab:paravar}
\end{table*}

\begin{table*}
\begin{center}
\begin{tabular}{cc|c|c|c|c|c|c|c|c|c|c|c|}
\cline{3-11}
& & \multicolumn{9}{ c| }{Training Data Percentage} \\ \cline{3-11}
& & \textbf{10\%} & \textbf{20\%} & \textbf{30\%} &\textbf{40\%} &\textbf{50\%} & \textbf{60\%} &\textbf{70\%} & \textbf{80\%} & \textbf{90\%} \\ \cline{1-11}
\multicolumn{1}{ |c }{\multirow{3}{*}{Metrics} } &
\multicolumn{1}{ |c| }{\textbf{Accuracy}} &0.674&0.672&0.683&\textbf{0.704}&0.697&0.698&0.669&0.673&0.656 \\ \cline{2-11}
\multicolumn{1}{ |c }{} &
\multicolumn{1}{ |c| }{\textbf{AUC}} &0.196&0.196&0.201&\textbf{0.224}&0.221&0.219&0.214&0.212&0.172 \\ \cline{2-11}
\multicolumn{1}{ |c }{} &
\multicolumn{1}{ |c| }{\textbf{F1 Measure}} &0.792&0.793&0.803&0.817&0.812&\textbf{0.818}&0.797&0.801&0.789 \\ \cline{1-11}
\end{tabular}
\end{center}
\caption{Performance of the framework with varying values of training size}
\label{tab:trainresults}
\end{table*}

\subsection{Effect of Varying Interaction Network Parameter}
In \textbf{Algorithm 1}, $w_{reg}$ represents the contributions of interaction network information to the framework. To understand the effect of this on the performance of the framework, we vary the value of the regularization parameter as $w_{reg}=$[0, 0.1, 0.2, 0.5, 1, 2, 5, 10]. We use the earliest 50\% of the data for training and the rest for testing. We study the variations of the performance of the framework for these values and illustrate the results in Table \ref{tab:paravar}.

From Table \ref{tab:paravar}, we can see that interaction network information is useful in improving the performance of the framework in all the three metrics. For very high values of the information network parameter ($ w_{reg} \geq 2 $) the performance slightly decreases, as this might suppress the contributions of the other terms. However, we can see that it still significantly outperforms the baselines for all values of the information network parameter. A paired t-test to compare the results with the baselines showed that the improvement is significant. Therefore, the framework is a robust to the variation of the information network parameter.

In summary, we can say that the information network helps in improving the overall performance of the framework, emphasizing the contribution of the regularization term in determining whether the next post of a user will be a declaration of protest. The framework performs consistently across all values of the parameter and hence is robust to its variation. We next study the effect of varying training data proportions on the performance of the framework.

\subsection{Variation of Training Data Size}
We now evaluate the variation of performance of the framework with different proportions of training data. This experiment enables us to assess the performance of the framework when less amount of training information is available. A possible scenario could be in the early stages of protest when only a few users are available for training. This experiment will also help assess the robustness of the framework for varying proportions of training data. We change the percentage of the training data set from 10\% to 90\% in steps of 10\%, starting from the earliest candidate post and measure the framework performance. We illustrate the results of the experiment in Table \ref{tab:trainresults} and make the following observations.

From Table \ref{tab:trainresults}, we can say that more training data is beneficial for increasing the performance of the framework. The framework outperforms the nearest baseline \cite{ma2014tagging}, for tiny proportions for training data (10\%), demonstrating that it performs well for low training data sizes. We also can see a slight dip in performance for higher proportions of training data ($> 80\%$), and this may be due to insufficient testing data. The performance shows consistent trends in all three metrics, showing that the framework efficiently utilizes the training data points to determine whether the next post of a user will be a declaration of protest.

In summary, the results demonstrate that the framework can learn from a small amount of training data, and it effectively utilizes training data to determine whether the next post of a user will be a declaration of protest. The framework consistently performs across all proportions of training data and hence is robust to its variations.

\section{Related Work}

The use of social media for political campaigns has received considerable attention in the literature \cite{segerberg2011social,gonzalez2011dynamics,ranganathunderstanding}. \cite{lee2013campaign} study the problem of extracting campaigns from social media using textual co-similarity, and the authors here do not concentrate on the behavior of individual users. The use of social media for political mobilization has been studied in \cite{bond201261}, and the authors analyze the effect of political messages on political self-expression and interactions. They, however, do not predict future declarations of protest of an individual user. Behavior adoption by social media users has been modeled as an effect of information propagation in \cite{myers2012clash,gomez2013modeling,lin2013bigbirds}. The effect of external influence on information propagation in networks have been studied in \cite{myers2012information,iwata2013discovering}. Propagation networks might not be fully available in many scenarios, and we utilize the user history to predict whether his next post will be a declaration of protest.

Considerable attention has been given to predicting characteristics of a user's future status messages from his history \cite{yang2012we,ranganathtimecritical,liang2012time,kumar2013whom}. The authors in \cite{godin2013using} predict the characteristics of the user's future messages by modeling the topics from his previous posts. This utilizes only the content of the individual user. The authors in \cite{kywe2012recommending} use collaborative filtering methods by incorporating the content of similar users and posts to compute characteristics of future messages. The authors of \cite{ma2014tagging} predict the characteristics of future messages by integrating the past content of the user, temporal information with the effect of interactions between candidate users. We build upon these works to model how messages interact with the user over time to affect the probability of his protest declaration in his next post.

The study of Brownian motion on networks has been investigated in \cite{zhou2003network,zhou2004network}. The theoretical development for adapting Brownian motion in the network environment has been considered in \cite{zhou2003network}, with applications in community discovery. It has also been used for model movements of stock price, due to its ability to model sharp changes \cite{tankov2003financial}. More recently, Brownian motion has been used to model information propagation processes in the presence of external influence in social networks \cite{jin2014modeling}. In most scenarios, the complete propagation network between the candidate users is not available. To address this, we take each user as a separate entity and model his past interactions and status messages to predict his protest participation.

\section{Conclusions and Future Work}
\label{sec:conclu}

Social media provides a platform for people to declare protest on various socio-political issues. In scenarios where the spread of protest is not available, predicting if the next post of a given user will be a declaration of protest will help in estimating the number of protest participants, facilitating better protest organization. In this paper, we draw from theories of protest participation and propose a framework that models the user's history to predict whether the next post of a given user will be a declaration of protest. We incorporate these principles to model the interplay between his content and interactions over time to predict latent dimension representations of the user's future post. We feed the latent dimensions to a classifier and determine whether the next post of a user will be a declaration of protest. We evaluate the framework on data posted in Twitter during protests on the Nigerian Elections. Our experiments demonstrate its effectiveness in predicting whether the next post of a user will be a declaration of protest.

This work paves the way to interesting future directions of research. Collecting labels of a user's posts over time can assist us in understanding the dynamical performance characteristics of the proposed model. Studying the behavior of protest groups over time will give insights into their formation and dynamics. The impact of the protest groups on spreading protest related messages is an interesting area of future research. Studying information seeking patterns among protesters can give interesting insights into their recruitment process. Finally, linguistic analysis of the messages posted during protests can provide insight into the persuasive content adopted by the protesters.

\section{Acknowledgements}
This material is based upon work supported by, or in part by, Office of Naval Research (ONR) under grant number N000141010091.

\bibliographystyle{aaai}
\bibliography{sigproc}

\end{document}